\documentclass[8.5pt,twoside,twocolumn]{article}
\usepackage[super,sort&compress,comma]{natbib} 
\usepackage{mhchem}
\usepackage{times,mathptmx}
\usepackage{sectsty}
\usepackage{balance} 
\usepackage{multirow}

\usepackage{graphicx} 
\usepackage{lastpage}
\usepackage[format=plain,justification=raggedright,singlelinecheck=false,font=small,labelfont=bf,labelsep=space]{caption} 
\usepackage{fancyhdr}
\begin{document}

\thispagestyle{plain}
\fancypagestyle{plain}{
\renewcommand{\headrulewidth}{1pt}}
\renewcommand{\thefootnote}{\fnsymbol{footnote}}
\renewcommand\footnoterule{\vspace*{1pt}%
\hrule width 3.4in height 0.4pt \vspace*{5pt}} 
\setcounter{secnumdepth}{5}

\makeatletter 
\def\subsubsection{\@startsection{subsubsection}{3}{10pt}{-1.25ex plus -1ex minus -.1ex}{0ex plus 0ex}{\normalsize\bf}} 
\def\paragraph{\@startsection{paragraph}{4}{10pt}{-1.25ex plus -1ex minus -.1ex}{0ex plus 0ex}{\normalsize\textit}} 
\renewcommand\@biblabel[1]{#1}            
\renewcommand\@makefntext[1]%
{\noindent\makebox[0pt][r]{\@thefnmark\,}#1}
\makeatother 
\renewcommand{\figurename}{\small{Fig.}~}
\sectionfont{\large}
\subsectionfont{\normalsize} 

\fancyfoot{}
\fancyfoot[RO]{\footnotesize{\sffamily{1--\pageref{LastPage} ~\textbar  \hspace{2pt}\thepage}}}
\fancyfoot[LE]{\footnotesize{\sffamily{\thepage~\textbar\hspace{2pt} 1--\pageref{LastPage}}}}
\fancyhead{}
\renewcommand{\headrulewidth}{1pt} 
\renewcommand{\footrulewidth}{1pt}
\setlength{\arrayrulewidth}{1pt}
\setlength{\columnsep}{6.5mm}
\setlength\bibsep{1pt}

\twocolumn[
  \begin{@twocolumnfalse}
\noindent\LARGE{\textbf{Orbital frustration in the $S = \frac{1}{2}$ kagome magnet vesignieite, BaCu$_{3}$V$_{2}$O$_{8}$(OH)$_{2}$}}
\vspace{0.6cm}

\noindent\large{\textbf{David Boldrin,\textit{$^{a}$} Kevin Knight\textit{$^{b}$} and Andrew S. Wills\textit{$^{a}$}$^{\ast}$}}\vspace{0.5cm}


\vspace{0.6cm}

\noindent \normalsize{}
\vspace{0.5cm}
 \end{@twocolumnfalse}
  ]

\section{Abstract}

Here we report crystallographic and magnetic studies on high quality samples of the magnetically frustrated $S = \frac{1}{2}$ kagome antiferromagnet vesignieite, BaCu$_{3}$V$_{2}$O$_{8}$(OD)$_{2}$. Powder neutron diffraction data collected from samples obtained by a new hydrothermal synthetic route reveal a previously unobserved trigonal $P3_{1}21$ structure, similar to the isoelectronic mineral SrCu$_{3}$V$_{2}$O$_{8}$(OH)$_{2}$. The refined structure is consistent with orbital frustration of the $e_{\mathrm{g}}$ $d$-orbitals in a sublattice of the Cu$^{2+}$ kagome network due to a dynamic Jahn-Teller effect, which persists below the magnetic transition at $T_{\mathrm{N}} = 9$\,K and makes the material an interesting candidate for exploring concomitant spin and orbital frustration. A combination of crystallographic strain analysis and magnetisation measurements indicate strong magnetostructural coupling which may explain the varied magnetic behaviour between samples of vesignieite in the literature. The revised orbital structure is similar to that found in volborthite, rather than the quantum spin liquid herbertsmithite, and provides a convincing argument for the differing magnetic properties found in these frustrated magnets.

\section{Introduction}
\footnotetext{\dag~Electronic Supplementary Information (ESI) available: [details of any supplementary information available should be included here]. See DOI: 10.1039/b000000x/}

\footnotetext{\textit{$^{a}$~Department of Chemistry, University College London, London, United Kingdom. E-mail: d.boldrin@ucl.ac.uk; a.s.wills@ucl.ac.uk} \textit{$^{b}$~ISIS Facility, Rutherford Appleton Laboratory, Chilton, Didcot, OX11 0QX, United Kingdom.}}

Spin-$\frac{1}{2}$ kagome magnets have attracted intense attention due to their ability to host exotic magnetism induced by geometric frustration. In classical systems, such as the $S > \frac{1}{2}$ jarosites and SCGO (SrCr$_{8}$Ga$_{4}$O$_{19}$) minerals, this often leads to spin liquid or spin glass properties \cite{Keren2000,Uemura1994,Yildirim2006,Wills2000,Wills2001,Bisson2008}. In the quantum limit, $S = \frac{1}{2}$, kagome magnets are the most promising model system to observe quantum spin liquids (QSLs) in 2-dimensions. These states are of tremendous interest as they could underpin a range of exotic electronic properties, such as the transition to high-temperature superconductivity in the doped cuprates \cite{Anderson1973,Anderson1987,Balents2010}. Whilst this attractive phenomena drove initial research in these materials, the richness of the $S = \frac{1}{2}$ kagome phase diagram has been revealed with the discovery of new experimental model materials:  the copper vanadate minerals vesignieite and volborthite display various magnetic phases from incommensurate order to heterogeneous static and dynamic spin states \cite{Ishikawa2015,Quilliam2011,Colman2011}; in paratacamite-based materials fractional excitations have been observed in herbertsmithite \cite{Han2012}, whilst kapellasite and haydeeite both lie in close proximity to a chiral QSL state \cite{Colman2010,Fak2012,Boldrin2015b,Iqbal2015,Bernu2013}. Beyond these well known materials, the recent discovery of multiferroicity in KCu$_{3}$As$_{2}$O$_{7}$(OD)$_{3}$ further exemplifies the range of behaviours possible in $S = \frac{1}{2}$ kagome magnets, whilst also highlighting their potential as functional materials \cite{Nilsen2014}.

Research into the detailed properties  of $S = \frac{1}{2}$ kagome magnets has so far concentrated on spin-only models and the consequences of orbital physics has largely been unconsidered. In contrast, the vast majority of experimental model materials are based on moment-bearing $d^{9}$ transition metals, which in octahedral geometries possess singly-occupied and doubly-degenerate $e_{\mathrm{g}}$ orbitals, and so are Jahn-Teller (JT) active. Such orbitally-active transition metal materials are often described by Kugel-Khomskii (KK)-type models, where orbital superexchange terms must be included to satisfactorily describe the observed physics \cite{Kugel1973}, and are exemplified by the material KCuF$_{3}$ \cite{Lee2011,Paolasini2002}. While these orbital and spin components often operate at drastically different energy scales, when they are of similar energies they may couple to form a modified Hamiltonian that necessarily has directional dependence following the anisotropy of orbital space \cite{Kugel1982,Tokura2000}. When combined with spin frustration and quantum effects, such as can arise in $S = \frac{1}{2}$ kagome magnets, the orbital frustration associated with orbital degeneracy is predicted to enhance quantum fluctuations and stabilise a quantum spin-orbital liquid (QSOL) \cite{Tokura2000,Li1998,Khomskii2003}. Recent experimental studies have begun to explore such ground states, such as in Ba$_{3}$CuSb$_{2}$O$_{9}$ \cite{Ishiguro2013} and LaSrVO$_{4}$ \cite{Dun2014} which show suppression of both orbital and spin ordering through geometric frustration. As of yet no model QSOL materials based on  $S = \frac{1}{2}$ kagome magnets have been studied and we argue that their ability to possess geometric and orbital frustration make them ideal templates to further investigate quantum spin-orbital frustration.

In this paper we re-evaluate the crystal structure of vesignieite following a new synthesic route to high-quality samples, and find it consistent with orbital frustration of the $e_{\mathrm{g}}$ $d$-orbitals of the Cu$^{2+}$ and a dynamic Jahn-Teller (JT) effect. This frustration occurs on a sublattice of the Cu kagome network and persists to $T = 4$\,K, well below the suppressed magnetic transition at 9\,K. A combination of crystallographic strain analysis and magnetic susceptibility measurements suggests a strong magnetostructural coupling, related to the complex JT effect, that may explain the varied magnetic behaviour observed between samples. Importantly, these results call for re-analysis of several Cu$^{2+}$ kagome magnets to include orbital physics and opens up several possibilities of exploring spin-orbital frustration in vesignieite and other related minerals.

\section{Experimental}

The new synthetic route for vesignieite was adapted from that used for `Sr-vesignieite' \cite{Boldrin2015}, whereby the layered copper vanadate mineral volborthite, Cu$_{3}$V$_{2}$O$_{7}$(OH)$_{2} \cdot 2$H$_{2}$O, is used as a starting reagent. Ba(CH$_{3}$CHOO)$_{2}$ (0.2167\,g, $8.467\times10^{-4}$\, mol) was placed in a PTFE-lined hydrothermal bomb (15\,ml) with volborthite (0.4018\,g, $8.467\times10^{-1}$) and 10\,ml H$_{2}$O (or D$_{2}$O for deuterated samples). The sealed bomb was heated from room temperature to 200\,$^{\circ}$C at a rate of 1\,$^{\circ}$C/min and left for 24\,hours before being removed from the furnace and left to cool naturally. The resulting green powder was collected and washed thrice with distilled water, then thrice with acetone, using centrifugation and dried on a rotary evaporator. To improve deuteration, samples were hydrothermally annealed at 200\,$^{\circ}$C for 24\,hours in D$_{2}$O (10\,ml). This was not found to impact the structure or crystallinity of the samples. Henceforth, the sample prepared under these conditions is referred to as sample 1.

For comparison with sample 1, additional samples were synthesised using the reflux method described previously in the literature \cite{Colman2011,Quilliam2011}. These were post-synthetically hydrothermally annealed in 10\,ml H$_{2}$O at 200\,$^{\circ}$C for between 0 and 168\,hours (7\,days) in PTFE-lined hydrothermal bombs. As such, samples prepared using this second method are referred to as 2-0, 2-2, 2-6, 2-24, 2-72 and 2-168 where the second number indicates the duration of hydrothermal annealing in hours.

Synchrotron powder diffraction data were collected at the high resolution 11-BM beamline at the APS, ANL. Powder neutron diffraction data was collected on 2.7\,g of sample 1 at the HRPD beamline, ISIS. Magnetic susceptibility measurements were collected on all samples using a MPMS-7 DC-SQUID.

\begin{figure}[h]
\centering
	\includegraphics[width=0.45\textwidth]{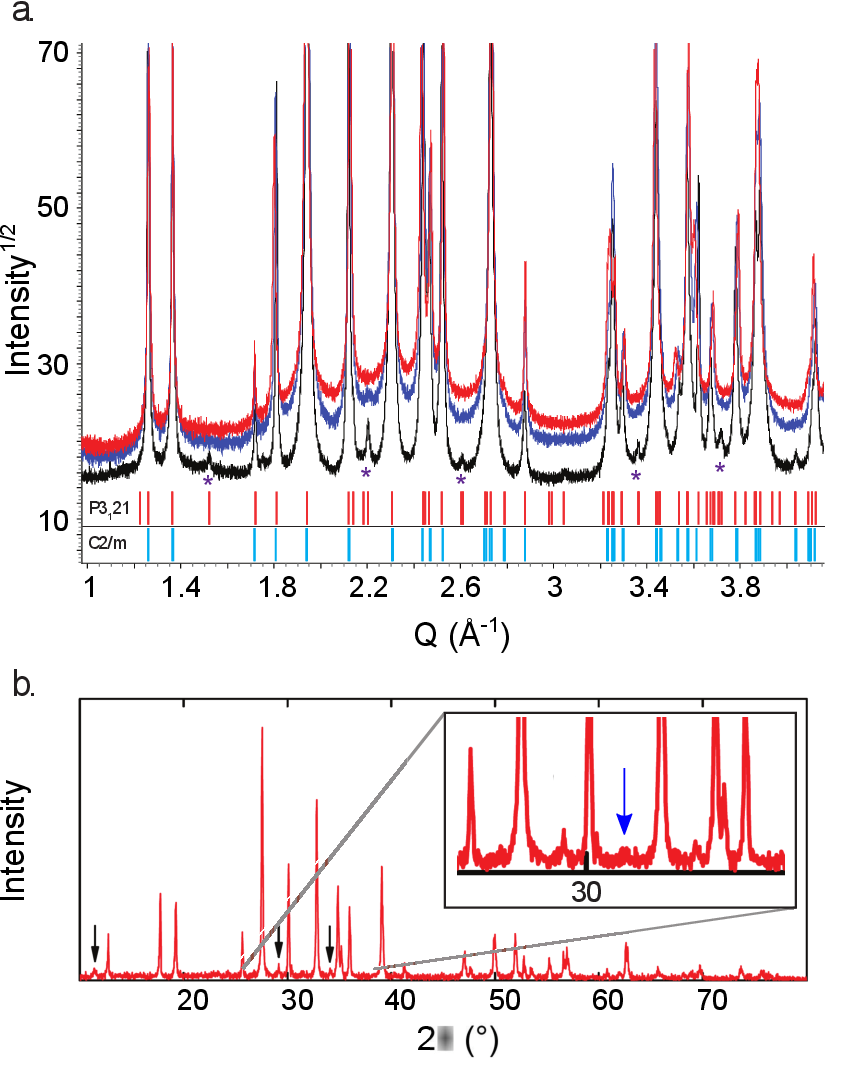}
  \caption{\textbf{a.} Synchrotron powder diffraction data collected from samples 1 (black), 2-0 (red) and 2-168 (blue). Several peaks are clearly visible in the data from sample 1 that cannot be indexed with $C12/m1$ symmetry, but instead $P3_{1}21$ as in `Sr-vesignieite'. Some of these peaks, indicated with asterisks, are also visible in the data from sample 2-168 but they are significantly broadened. \textbf{b.} X-ray diffraction data taken from \cite{Yoshida2013}. The peak at $\sim 35$\,$^{\circ}$, indicated in the inset, aligns with the most intense additional peak of the $P3_{1}21$ structure at $Q \sim 2.2$\,\AA$^{-1}$ in \textbf{a}.}
  \label{fgr:example}
\end{figure}

\section{Results}

\subsection{Powder synchrotron diffraction}

Powder synchrotron diffraction data were collected from all samples over d-spacings between 0.5 and 7\,\AA$^{-1}$ for a total of 30 minutes at $T = 300$\,K. Samples were packed in 0.8\,mm diameter Kapton tubes, giving a $\mu R = $0.76. Although this value is relatively low, Rietveld refinement of the oxygen positions for all datasets proved unstable, likely due to the presence of several heavy elements, and therefore a full structural refinement was impossible. Consequently, we limit our discussion to Pawley refinements and crystallite strain analysis. These analyses were performed using the TOPAS software package \cite{TOPAS}.

A comparison of the synchrotron data collected from samples 1, 2-0 and 2-168 is shown in Figure 1a. The diffraction pattern of sample 1 has several peaks that cannot be indexed with the $C12/m1$ monoclinic structure previously reported for vesignieite \cite{Zhesheng1991}, most notably at $Q \sim 2.2$\,\AA$^{-1}$. Instead the pattern indexes with the trigonal $P3_{1}21$ structure recently reported for `Sr-vesignieite' and a Pawley refinement starting with the unit cell from the latter material fits the data excellently and accounts for all observed peaks \cite{Boldrin2015}. Data collected on the samples 2-0 and 2-168 have a noticeably higher background and many of the peaks associated with the $P3_{1}21$ structure are not visible. However, the peak at $Q \sim 2.2$\,\AA$^{-1}$ is visible in the spectra of sample 2-168, although it is significantly broadened, particularly with respect to the rest of the pattern. Figure 1b shows a powder X-ray diffraction spectrum taken on a hydrothermally annealed sample made by Yoshida \emph{et al.} and this peak, at $\sim 35$\,$^{\circ}$($2\theta$), is again visible \cite{Yoshida2013}. In that study, the observation of a crystalline impurity meant that small features were either ignored or assigned to the impurity phase, thus the monoclinic structure was assumed to be correct and no re-analysis of the structure was performed. Unfortunately, this structure cannot be reconciled with that of $\beta$-vesignieite where rhombohedral $R\overline{3}m$ symmetry is determined from single crystal diffraction \cite{Yoshida2012}. We note that these crystals were synthesised hydrothermally, similarly to samples 2-0 through 2-168 here, and therefore the additional peaks may have been broad and again not indexed in that study.

\begin{figure}[!ht]
\centering
	\includegraphics[width=0.4\textwidth]{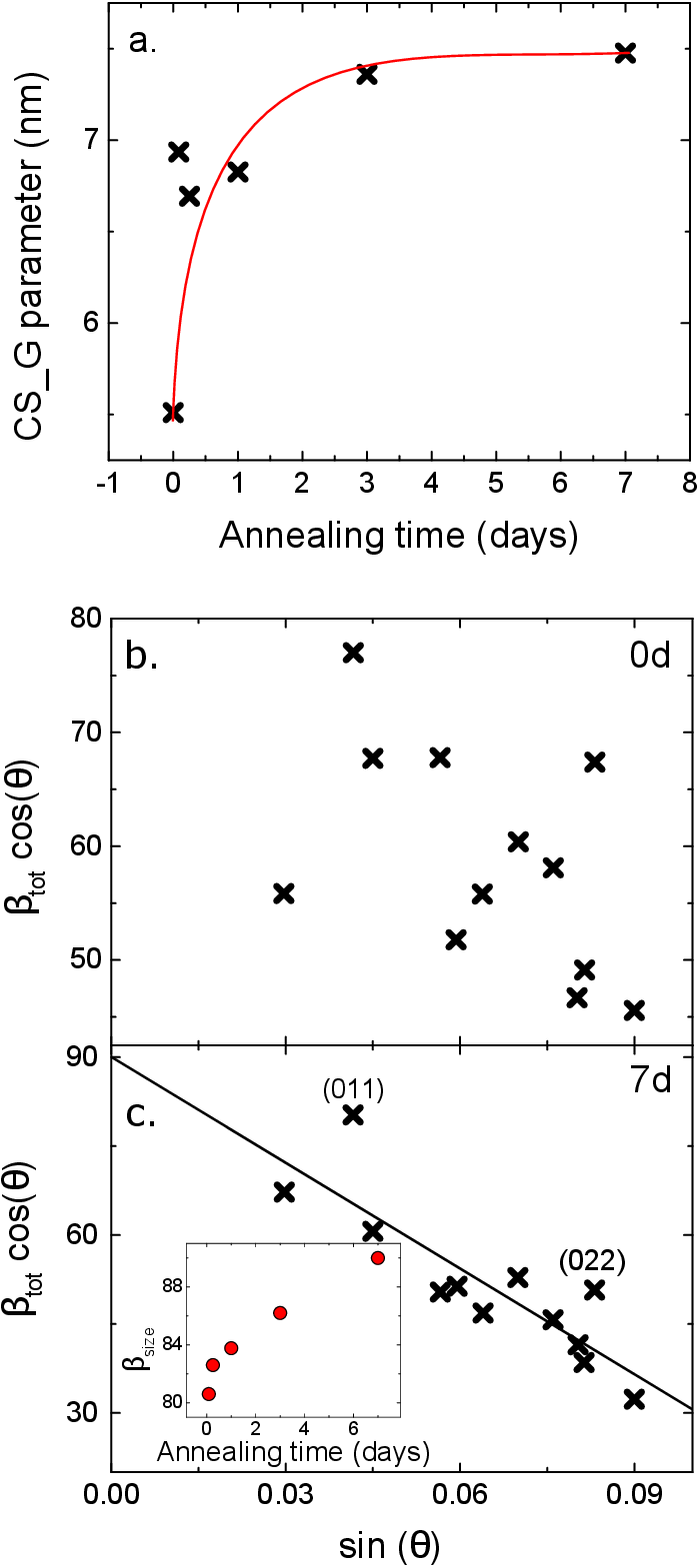}
  \caption{\textbf{a.} A plot of the Gaussian profile parameter, CS\_G in TOPAS, for the trigonal $P3_{1}21$ phase as a function of annealing time. \textbf{b-c.} Williamson-Hall plots built from the synchrotron powder diffraction data collected on Samples 2-0 (\textbf{b.}) and 2-168 (\textbf{c.}). The inset of the lower figure shows the variation of the fitted crystallie size, $\beta_{\mathrm{size}}$ as a function of annealing time.}
  \label{}
\end{figure}

Pawley refinement of the data collected from sample 2-168 with a $P3_{1}21$ structure gives a particularly poor fit to the additional peaks not present in monoclinic symmetry as they are significantly broadened. To achieve a good fit, a two-phase refinement model was used: one phase with monoclinic $C12/m1$ symmetry and another with trigonal $P3_{1}21$. The profile of both phases were independently modelled with a simple convolution of Gaussian and Lorentzian peakshapes, parameters CS\_G and CS\_L in TOPAS respectively, the values of which correlate with particle size. These profile parameters were allowed to refine freely in the Pawley refinements and the same procedure was used for data collected on samples 2-0, 2-2, 2-6, 2-24 and 2-72.

\begin{table*}[t]%
\small
\centering
	\begin{tabular}{| c | c | c | c | c | c | c | c |}
	\hline
	 & & \multicolumn{2}{| c |}{Cu1} & \multicolumn{2}{| c |}{Cu2} & \multicolumn{2}{| c |}{V} \\
	\hline
	 & & Bond length (\AA) & Bond difference (\AA) & Bond length (\AA) & Bond difference (\AA) & Bond length (\AA) & Bond difference (\AA) \\
	\hline
	\hline
	\multirow{2}{*}{O11} & long & 2.15 (4) & \multirow{2}{*}{0.08 (7)} & 2.29 (4) & \multirow{2}{*}{0.29 (6)} & 1.53 (5) & \multirow{2}{*}{0.29 (6)} \\
	 & short & 2.07 (6) &  & 2.00 (5) & & 1.82 (4) & \\
	\hline
	\multirow{4}{*}{O12} & long & 2.51 (3) & \multirow{2}{*}{0.32 (7)} & & & \multirow{2}{*}{1.91 (2)} & \multirow{4}{*}{0.21 (2)} \\
	 & short & 2.20 (6) & & & & & \\
	 & long & 2.58 (3) & \multirow{2}{*}{0.36 (7)} & & & \multirow{2}{*}{1.70 (1)} & \\
	 & short & 2.22 (6) & & & & & \\
	\hline
	\multirow{2}{*}{O13} & long & 2.12 (3) & \multirow{2}{*}{0.13 (4)} & 2.42 (2) & \multirow{2}{*}{0.38 (2)} & 1.86 (2) & \multirow{2}{*}{0.18 (5)} \\
	 & short & 1.99 (3) &  & 2.03 (2) & & 1.69 (1) & \\
	\hline
	\multirow{2}{*}{O21(H)} & & 1.91 (2) & & \multirow{2}{*}{1.86 (3)} & & \multirow{2}{*}{1.57 (1)} & \\
	 & & 1.93 (2) & & & & & \\
	\hline
	\end{tabular}
	\caption{Refined Cu\---O and V\---O bond lengths from the powder neutron diffraction data collected at $T = 4$\,K - the crystal structure is shown in Figure 3b. Multiple bond lengths are given for certain oxygen sites because these were split for the refinement to model the positional disorder. The large bond difference for the Cu2\---O11 and Cu2\---O13 bonds, combined with the short Cu2\---O21(H) bond, is consistent with a dynamic JT effect.}
	\label{}	
\end{table*}

A plot of the Gaussian profile parameter for the $P3_{1}21$ phase, CS\_G, as a function of annealing time is shown in Figure 2a. The larger CS\_G value with longer annealing time is a result of the sharper profile function for the $P3_{1}21$ phase, hence the additional peaks visible in the data collected on sample 2-168. An important feature of the Pawley refinements is that even for sample 2-0 where the $P3_{1}21$ phase profile is broad, the two phase model significantly improves the fit in the regions of the additional peaks due to the large background. Evidently this anisotropic profile broadening is strictly sample related and is commonly caused by microstructural properties such as defects, dislocations and stacking faults \cite{Stephens1999}.

To further characterise the structural disorder, size and strain properties were determined from Williamson-Hall plots \cite{powderdiffraction}. This method assumes that the sample broadening of each peak associated with a reflection is a simple convolution of crystallite size and lattice strain, where both vary differently as a function of $2\theta$. Accordingly, the total internal breadth of each peak, $\beta_{\mathrm{tot}}$, is given by:

\begin{equation}
\beta_{\mathrm{tot}} = \beta_{L} + \beta_{\epsilon} = \epsilon\ \mathrm{tan}\theta + \frac{1}{L\ \mathrm{cos}\theta}
\label{eq:}
\end{equation}

\noindent
where $\beta_{\epsilon}$ and $\beta_{L}$ are the internal breadths of the size and strain broadening, $\beta$ and $\epsilon$, respectively. Multiplying this equation by cos$\theta$ then gives:

\begin{equation}
\beta_{\mathrm{tot}}\ \mathrm{cos}\theta = \epsilon\ \mathrm{sin}\theta + \frac{1}{L}
\label{eq:}
\end{equation}

\noindent
so that a plot of $\beta_{\mathrm{tot}}$\,cos\,$\theta$ \emph{vs.} sin\,$\theta$ should give a straight line with $L$ as the $y$-intercept and $\epsilon$ as the gradient. Williamson-Hall plots constructed from the synchrotron data collected on Samples 2-0 and 2-168 are shown in Figure 2b. $\beta_{\mathrm{tot}}$ for Sample 2-0 appears to have no relation to $\theta$, rendering a straight-line fit impossible and suggesting significant anisotropic strain and size effects. After hydrothermal annealing for 7\,days, the majority of $\beta_{\mathrm{tot}}$ values fall approximately onto a straight line, except for those associated with the $(0kk)$ reflections. These lie above the fit indicating that they are more strained. Taking into account all reflections, the fitted size component, $\beta_{\mathrm{size}}$, increases as a function of annealing time (Figure 2b inset) showing an improved sample crystallinity. The strain also increases although this may be due to compression from lattice shrinkage as evidenced from the refined unit cell volume (not shown), a phenomena also found in Williamson-Hall analysis of other materials \cite{KhorsandZak2011}.

\begin{figure}[!hb]
\centering
	\includegraphics[width=0.48\textwidth]{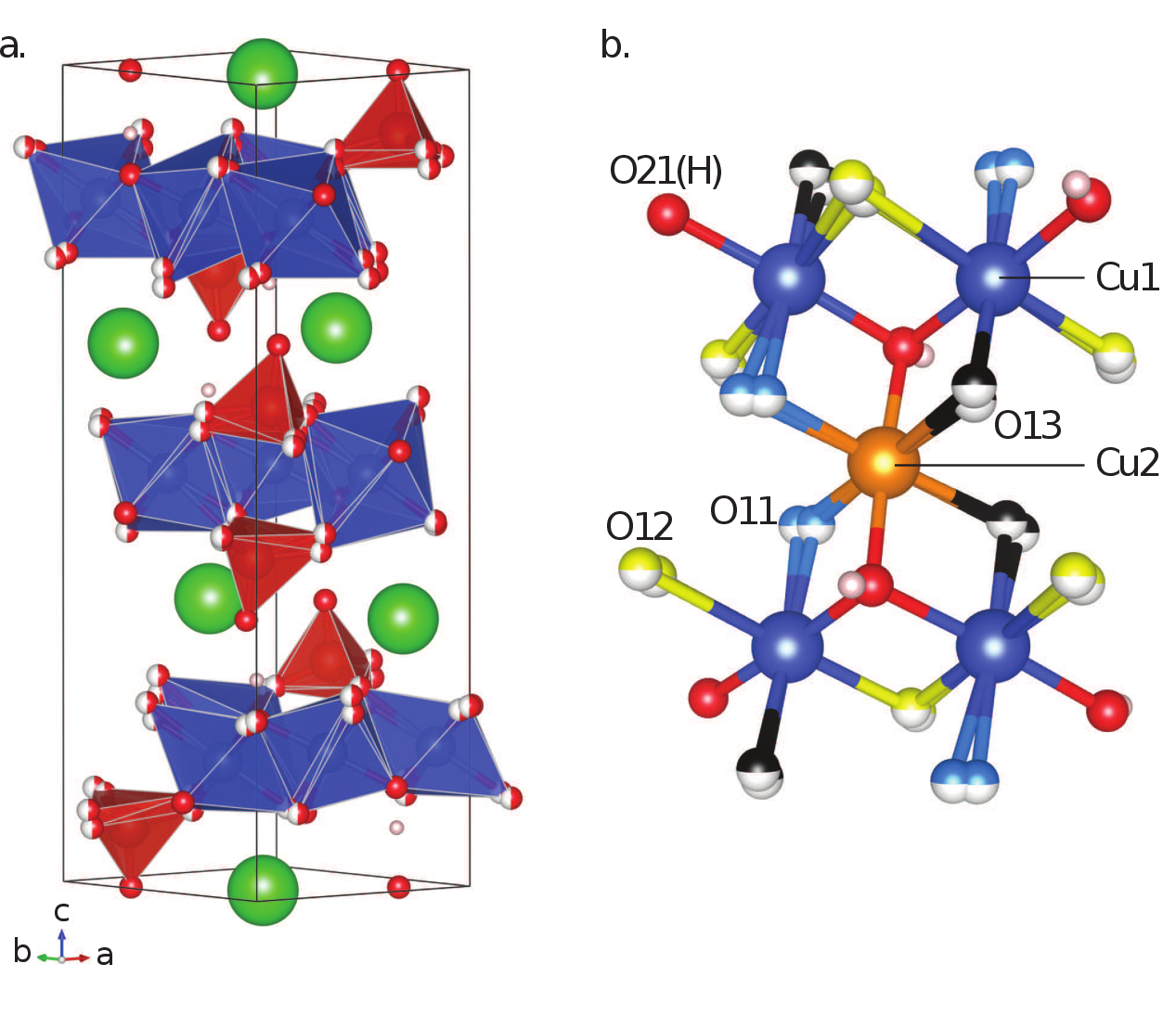}
  \caption{\textbf{a.} The refined crystal structure of vesignieite at $T = 4$\,K obtained from Rietveld refinement of powder neutron diffraction data. The final Rietveld refined profiles are shown in Figure S1. \textbf{b.} The local coordination of the two Cu sites and associated Cu\---O bond lengths. The orientation of the split O11 and O13 sites are approximately perpendicular to the bonds with the Cu1 site and parallel to those with Cu2. These orientations, along with the bond length variation observed in the refined bond lengths (Table 1) is consistent with static and dynamic Jahn-Teller order on the Cu1 and Cu2 sites, respectively.}
  \label{}
\end{figure}

\subsection{Powder neutron diffraction}

To obtain accurate atomic positions for the $P3_{1}21$ structure of Sample 1, powder neutron diffraction data were collected at the high-resolution powder diffraction (HRPD) beamline at ISIS, UK. 2.7\,g of sample was loaded in a standard rectangular geometry vanadium holder of 10\,mm depth. Data were collected at 4 and 20\,K using a standard orange cryostat for durations of 7\,h at each temperature. Rietveld refinement of the data was performed using the TOPAS software package \cite{TOPAS}, with a starting atomic model adapted from `Sr-vesignieite' \cite{Boldrin2015} but with lattice parameters taken from the Pawley refinements of the synchrotron data discussed previously. The final refinement of the data collected at 4\,K achieved goodness-of-fit parameter $R_{\mathrm{wp}} = 2.463$ .

To further characterise the JT states of the Cu$^{2+}$ sites, refinement of anisotropic displacement parameters on the octahedral oxygen sites was attempted, however these proved unstable and so alternative refinement techniques were required. The most stable approach involved first performing a thorough Rietveld analysis whereby the positions and isotropic displacements of all atomic positions were refined, bar the vanadium which was fixed to that from the `Sr-vesignieite' structure. Once convergence was achieved, the oxygen sites of the CuO$_{6}$ octahedra were split, and set to half occupancy, with the position of each allowed to refine freely and the isotropic displacements of all split oxygens fixed to be equal. For the data collected at $T = 4$\,K, this procedure lowered the $R_{\mathrm{wp}}$ and $\chi^{2}$ values from 3.732 to 3.617 and 2.372 to 2.274, respectively. 

The refined crystal structure of vesignieite at $T = 4$\,K, shown in Figure 3a with refined structural parameters in Table S1, closely resembles that of `Sr-vesignieite'. The larger Ba$^{2+}$ ion in vesignieite causes an enlargement of the unit cell ($a = 5.91785(4)$ and $c = 20.7599(4)$\,\AA) compared to its Sr$^{2+}$-analogue. Moreover, vesignieite shows the same pattern of axially compressed and rhombically distorted CuO$_{6}$ octahedra that has been found in `Sr-vesignieite', as shown in Figure 3b. 

Looking at the axially-compressed Cu2 site, we note firstly that this geometry is surprisingly rare, and that the observation of axially compressed Cu$^{2+}$ by powder diffraction can be explained in the vast majority of cases by a dynamic JT state involving the re-orientation of an \emph{axially elongated} coordination geometry\cite{Burns1996}. In vesignieite the split O11 and O13 sites are approximately aligned parallel to the Cu2\---O11/O13 bonds (Figure 3b). These O11 and O13 sites are shared between neighbouring Cu1 and Cu2 octahedra, however they are arranged approximately perpendicular to the Cu1\---O11/O13 bonds. Evidence for this is found in the difference between the Cu\---O bond lengths of the split oxygen sites: the Cu2\---O11/O13 bonds differ by 0.29(6)\,\AA\ ($\sim 13$\,\%) and 0.38(4)\,\AA\ ($\sim 17$\,\%), respectively, whilst the Cu1\---O11/O13 differ by significantly less: 0.08(7)\,\AA\ ($\sim 4$\,\%) and 0.13(4)\,\AA\ ($\sim 6$\,\%), respectively (Table 1). Such a large difference between the refined Cu2\---O bond lengths is consistent with fluctuations between two elongated Jahn-Teller configurations aligned along either pair of Cu2\---O11/O13 bonds, \emph{i.e.} a dynamic JT effect. 

Turning attention to the rhombically distorted octahedra of the Cu1 site, it is noteworthy that the Cu1\---O12 bonds also vary greatly (0.32(7)\,\AA, $\sim 14$\,\%) despite the small difference found in the Cu1\---O11 and Cu1\---O13 bonds mentioned previously. For a dynamic Jahn-Teller effect one expects bond elongation along two or more axes of the octahedron, therefore this appears an unlikely cause. Instead, we propose that it arises from the fluctuations on the Cu1 site, which due to the close connectivity of the two Cu sites cause rotation of the Cu1O$_{6}$ octahedra and, subsequently, the large observed difference in Cu1\---O12 bond lengths. Therefore, we conclude that the Cu1O$_{6}$ octahedra is most likely in a static JT state at $T = 4$\,K.

\subsection{SQUID}

\begin{figure}[]
\centering
	\includegraphics[width=0.45\textwidth]{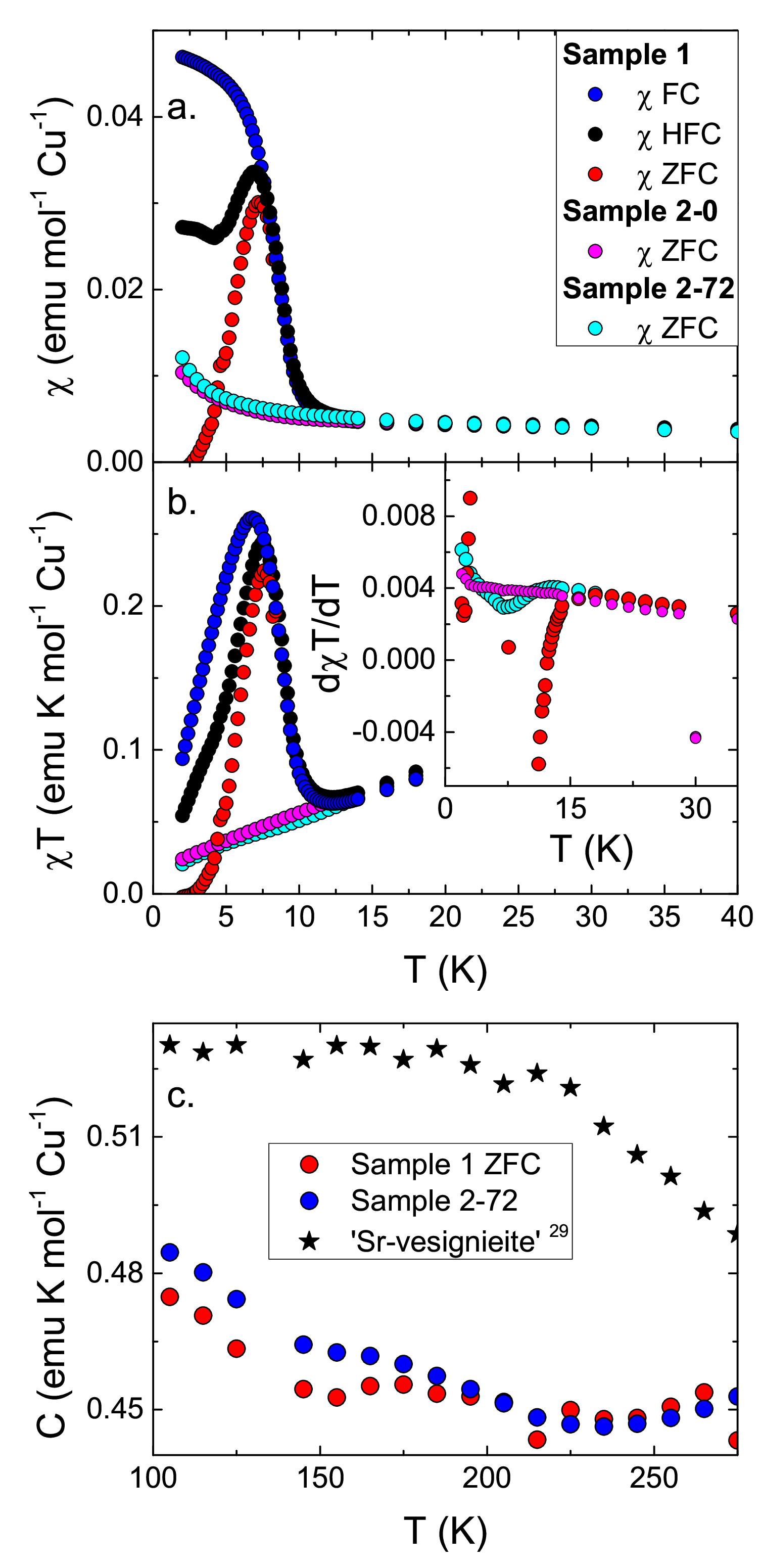}
  \caption{ \textbf{a.} A plot of magnetic susceptibility, $\chi$, as a function of temperature for samples 1, 2-0 and 2-72 using a measuring field of $H = 100$\,Oe in zero field cooled (ZFC); sample 1 under field cooled (FC) and high field cooled (HFC) conditions is also shown. \textbf{b.} $\chi T$ values for the same measurements on the same samples. The insets of both figures show the first derivatives of their respective data. \textbf{c.} The fitted Curie constant, $C$, as a function of temperature for samples 1, 2-72 and `Sr-vesignieite' \cite{Boldrin2015}. Both vesignieite samples have significantly reduced $C$ in this temperature range suggesting all Cu sites remain dynamically distorted above $T \sim 100$\,K.}
  \label{SQUID1}
\end{figure}

The low temperature magnetic susceptibility, $\chi$, of samples 1, 2-0 and 2-72 is shown in Figure \ref{SQUID1}a. Data were all collected on warming from $T = 2$ to 300\,K in an applied field of 100\,Oe after initially cooling in zero-field (ZFC);  for sample 1 data were also taken after cooling in the measuring field (FC) and an applied field of 70000\,Oe (HFC). For the moment, we will concentrate on the data collected on sample 1. At higher temperatures than those shown (Figure S2) $\chi$ approximates to the Curie-Weiss law, with a Weiss temperature of $\theta_{\mathrm{W}} = -75$\,K in good agreement with other vesignieite samples \cite{Colman2011,Yoshida2013}. A clear transition occurs at $T_{\mathrm{N}} \sim 9$\,K that is remarkably similar to that previously observed in both $\beta-$vesignieite and vesignieite samples hydrothermally annealed at very high temperatures \cite{Yoshida2012,Yoshida2013}. Whilst the transition resembles a ferromagnetic ordering, re-plotting $\chi$ \emph{vs.} $T$ as $\chi T$ \emph{vs.} $T$ (Figure 4b) shows both a sharp increase at $T_{\mathrm{N}} \sim 9$\,K followed by a comparably sharp decrease below $T \sim 7$\,K suggestive of both ferromagnetic and antiferromagnetic components to the ordering. The presence of a ferromagnetic component is confirmed by hysteresis in magnetisation measurements as a function of field at $T = 2$\,K with a small spontaneous moment of 0.02\,$\mu_{\mathrm{B}}$ (Figure S3, S4). 

While no clear transition is seen in the temperature-dependence of $\chi$ or $\chi T$ for samples 2-0 or 2-72 at any temperature (Figure 4a and b), the derivative of $\chi T$ (Figure 4b inset) suggests that there is a weak transition in the 2-72 sample at a similar temperature to that of sample 1. We speculate that the increase in the strength of this transition follows from the reduction in crystallographic disorder that was found in the refinements of the synchrotron data, and that these samples feature similar underlying magnetic responses, allbeit with a magnitude that depends on the level of disorder present.

In our study of `Sr-vesignieite', the temperature dependence of the fitted Weiss temperature, $\theta_{\mathrm{W}}$, reveals a subtle change of the magnetic response at $T_{\mathrm{JT}} \sim 230$\,K that indicates the onset of static Jahn-Teller order on some of the Cu$^{2+}$ sites\cite{Boldrin2015}. A similar analysis is presented in Figure 4c for Samples 1 and 2-72 for the calculated Curie constant, whereby $C$ is determined from Curie-Weiss fits to $\chi^{-1}(T)$ in 50\,K regions and plotted as a function of temperature. Both samples 1 and 2-72 exhibit a similar response, with $C$ increasing gradually between $100 < T < 300$\,K. Interestingly, the fitted $C$ values are significantly less than that of the isoelectronic `Sr-vesignieite'. Defining $C$ from the Curie law as

\begin{equation}
C = \frac{\mu_{0}\mu_{\mathrm{B}}^{2}}{3k_{\mathrm{B}}}Ng^{2}J(J+1)
\label{eq:}
\end{equation}

\noindent
one would naively expect it to be of similar value in the two materials and the discrepancy must originate from a modification of $J$ or $g$. Such an effect may occur due to changes in the orbital contributions to the magnetism of Cu$^{2+}$ and, in fact, this is often the case for Cu$^{2+}$ containing compounds, as discussed previously. Therefore, we suggest that the onset of static JT order which occurs in `Sr-vesignieite', and causes the increase of $C$ below $T = 300$\,K, is shifted to lower temperatures in the vesignieite samples presented. Unfortunately, deviations from Curie-Weiss behaviour below $T = 100$\,K means that the evolution of $C$ to lower temperatures cannot be determined. As the value of $C$ at $T = 100$\,K in vesigneite is still increasing towards that of `Sr-vesignieite', and in the orbitally ordered regime of both materials we may expect them to be of similar value, we conclude that the onset of static JT order is not complete in the temperature window studied.

\section{Discussion}

\begin{figure}[!ht]
\centering
	\includegraphics[width=0.50\textwidth]{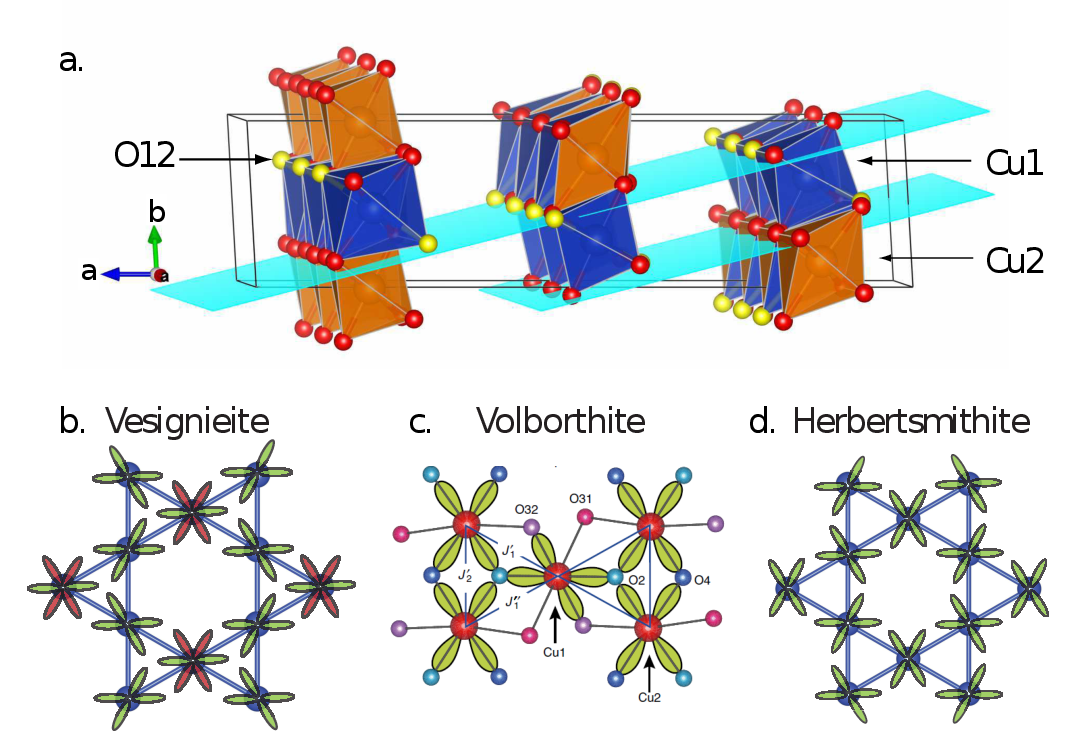}
  \caption{\textbf{a.} The (011) and (022) lattice planes of the vesignieite structure. These largely encompass the oxygens shared between the Cu sites, in particular the O12 sites (yellow) shared between the Cu1 (blue) site. The strain in the $(0kk)$ planes seen in the powder synchrotron diffraction data is likely a consequence of the transition from dynamic to static Jahn-Teller order along the Cu1 chains. \textbf{b.-d.} Orbital ordering diagrams of vesignieite and two other model $S = \frac{1}{2}$ KAFMs volborthite (take from \cite{Yoshida2012b}) and herbertsmithite. The superexchange mediating $d_{x^{2}-y^{2}}$ orbitals of vesignieite and volborthite are not related by 3-fold rotational symmetry whilst those of herbertsmithite are.}
  \label{Discussion1}
\end{figure}

Our crystallographic characterisation of new, high-quality vesignieite samples show that its structure is well described by the trigonal $P3_{1}21$ structure proposed for  `Sr-vesignieite'. Evidence for this structure is also found in poorer quality samples, although the associated peaks are significantly broadened suggesting microstructural strains and/or defects. Hydrothermal annealing induces crystallographic ordering of the $P3_{1}21$ structure, as evidenced from the synchrotron data, crystallographic size/strain analysis and comparisons with data already in the literature. In contrast, our new synthetic method crystallises  with a fully ordered structure. 

Close inspection of the Williamson-Hall plots (Figure 2) shows that the $0kk$ reflections in the data from sample 2-168 lie significantly further from the straight line fit than other crystallographic directions (Figure 2c). Interestingly, the $(0kk)$ planes, shown in Figure 5a, encompass several of the split oxygen sites and in particular the O12 site, which is shared between neighbouring Cu1 sites along the Cu1 chains. This observation hints at a relationship between the crystallographic strain and the static $\leftrightarrow$ dynamic JT transition on the Cu1 site, an effect commonly observed in other Jahn-Teller active materials \cite{Gehring1975}. Although the mechanism behind the reduction in strain by post-synthetic hydrothermal annealing remains unclear, we suggest that it is likely to be related to the complex JT effects of these materials.

The varied behaviour of the magnetic susceptibility seen in the vesignieite samples is further evidence of magnetostructural effects. The improved samples, available by hydrothermal annealing or the new synthetic route presented here, show a  considerably strengthened magnetic transition at $T_{\mathrm{N}} \sim 9$\,K that closely resembles that expected for a long-range ordered antiferromagnet \cite{Yoshida2012,Yoshida2013}. In poorer quality samples the transition at $T_{\mathrm{N}}$ has been shown to be inhomogeneous with both fluctuating and static spins \cite{Okamoto2009,Colman2011,Quilliam2011}.

It is difficult to say without further study how orbital frustration affects the magnetic response of vesignieite and whether it underlies discrepancies seen between samples and synthetic methods. In the case of the spin- and orbitally-frustrated LaSrVO$_{4}$, orbital fluctuations `melt' the magnetic order, suppressing the magnetic transition well below that predicted \cite{Dun2014}. In contrast, another common observation is that orbital ordering occurs concurrently at magnetic transitions \cite{Pen1997}, such as in LiVO$_{2}$ where the superexchange becomes strongly non-uniform at the magnetic transition and changes the Heisenberg approximation \cite{Vernay2004,Reynaud2001}.  Returning to vesignieite, we note that a dynamic JT effect may also explain the high-$T$ dependence of the ESR linewidth and high $g$-factors, which cannot be explained by a spin-only model\cite{Zorko2013}. Instead, the behaviour closely resembles that recently found in Ba$_{3}$CuSb$_{2}$O$_{9}$ which possesses a dynamic JT effect and short-range spin-orbital correlations \cite{Do2014,Zhang2010}. Therefore, further experiments are essential to better characterise the coupling between the orbital and spin degrees-of-freedom in vesignieite, such as whether spin and orbital order occur concomitantly or if a portion of the spins remain dynamic below $T_{\mathrm{N}}$ in these new samples.

As a model $S = \frac{1}{2}$ kagome magnet, it is important to now discuss the revised superexchange pathways within the revised trigonal crystal structure of vesignieite. As the CuO$_{6}$ octahedra are axially elongated, whether static or dynamic, the lone electron resides in the d$_{x^{2}-y^{2}}$ orbital, rather than the d$_{z^{2}}$ as previously proposed \cite{Okamoto2009}. Thus, as in `Sr-vesignieite', when the Cu1 sites undergo a transition to static JT order the chains will be antiferro-orbitally ordered, with the orbitally disordered Cu2 sites completing the kagome. A schematic of this orbital ordering pattern is shown in Figure 5, alongside the other $S = \frac{1}{2}$ kagome magnet models volborthite and herbertsmithite. Vesignieite shares a similar orbital ordering to volborthite: both have the $d_{x^{2}-y^{2}}$ orbitals arranged around the kagome triangles such that there is no 3-fold rotational symmetry relating them and the orbital superexchange paths are inequivalent. Where the $d_{z^{2}}$ orbitals mediate superexchange for all the Cu$^{2+}$ ions, these paths would be equivalent. Contrastingly, herbertsmithite has fully ordered $d_{x^{2}-y^{2}}$ orbitals that are related by 3-fold rotational symmetry at the centre of each kagome triangle, such that all superexchange paths are equivalent. No magnetic transition is observed in this latter material down to $T = 20$\,mK \cite{Hiroi2009}. As such, the symmetry of the orbitals, and therefore that of the symmetric and antisymmetric superexchange contributions, provides the most convincing reason behind the contrasting magnetic behaviour in these $S = \frac{1}{2}$ kagome magnets.

\section{Conclusion}

We have prepared and characterised samples of the $S = \frac{1}{2}$ kagome magnet vesignieite and determined its true crystallographic structure. The refined structure is consistent with orbital frustration due to a dynamic JT effect on a sublattice of the Cu$^{2+}$ kagome network that persists below the magnetic transition at $T_{\mathrm{N}} = 9$\,K. Whilst the magnetic behaviour below $T_{\mathrm{N}}$ is highly sample dependent, the intimate connection between microstructural defects due to the dynamic JT effect and the observed magnetism appears to underlie these previously unexplained discrepancies. Our analysis also revises the previously proposed orbital ordering pattern so that the superexchange paths connecting the Cu$^{2+}$ are qualitatively distinct, whereas previously they were though to be essentially equivalent. The proposed orbital structure is remarkably similar to volborthite, rather than herbertsmithite, and provides the most convincing argument as to the different magnetic properties observed in these three model $S = \frac{1}{2}$ kagome magnets. Our findings call for further work on vesignieite and related Cu$^{2+}$-based kagome magnets, with an emphasis on the orbital structure, as they provide interesting models to study concomitant spin and orbital frustration.

\section{Acknowledgements}

D.B. would like to thank University College London for studentship funding. This research used resources of the Advanced Photon Source, a U.S. Department of Energy (DOE) Office of Science User Facility operated for the DOE Office of Science by Argonne National Laboratory under Contract No. DE-AC02-06CH11357. Experiment at the ISIS neutron source was supported by a beam-time allocation from the Science and Technology Facilities Council (STFC).

\footnotesize{
\bibliographystyle{rsc}

\providecommand*{\mcitethebibliography}{\thebibliography}
\csname @ifundefined\endcsname{endmcitethebibliography}
{\let\endmcitethebibliography\endthebibliography}{}

}

\end{document}